\documentclass[twocolumn,amsmath,amssymb, apl]{revtex4-1}
\usepackage{graphicx}    % Include figure files \
\usepackage{bm}% bold math
\usepackage{hyperref}    % add hypertext capabilities
\usepackage[mathlines]{lineno} % Enable numbering of text and display
\usepackage{xcolor}
\usepackage[percent]{overpic}
\usepackage{setspace}
\begin{document}

%\preprint{APS/123-QED}

\title{Force sensing with an optically levitated charged nanoparticle}

\author{David Hempston, Jamie Vovrosh, Marko Toro\v{s}, George Winstone}
\author{Muddassar Rashid}
\email{m.rashid@soton.ac.uk}
\author{Hendrik Ulbricht}
\email{h.ulbricht@soton.ac.uk}
\affiliation{
  Department of Physics and Astronomy, University of Southampton, SO17
  1BJ, United Kingdom}

\date{\today}

\begin{abstract} Levitated optomechanics is showing potential for
  precise force measurements.  Here, we report a case study, to show
  experimentally the capacity of such a force sensor. Using an
  electric field as a tool to detect a Coulomb force applied onto a
  levitated nanosphere. We experimentally observe the spatial
  displacement of up to 6.6 nm of the levitated nanosphere by imposing
  a DC field. We further apply an AC field and demonstrate resonant
  enhancement of force sensing when a driving frequency,
  $\omega_{AC}$, and the frequency of the levitated mechanical
  oscillator, $\omega_0$, converge. We directly measure a force of
  $3.0 \pm 1.5 \times 10^{-20}$ N with 10 second integration time, at
  a centre of mass temperature of 3 K and at a pressure of
  $1.6 \times 10^{-5}$ mbar.
\end{abstract}

\pacs{Valid PACS appear here}
\maketitle
% --------------------------------------------------
%             Introduction
% --------------------------------------------------
The ability to detect forces with increasing sensitivity, is of
paramount importance for many fields of study, from detecting
gravitational waves \cite{Abbott2016} to molecular force microscopy of
cell structures and their dynamics \cite{Goktas2017}. In the case of a
mechanical oscillator, the force sensitivity limit arises from the
classical thermal noise, as given by,
\begin{equation} \label{eq:thforce}
  S_{FF}^{th}= \sqrt{4k_b T m\omega_0/Q_m}.
\end{equation}
Where, $k_b$ is the Boltzmann constant, $T$ is the temperature of the
thermal environment, $m$, the mass of the object, $\omega_0$ is the
oscillator angular frequency, $Q_m = \omega_0/\Gamma_0$ is the
mechanical quality factor and $\Gamma_0$ is the damping factor. In
recent decades, systems, such as cold-atoms traps, have pushed the
boundaries of force sensitivities down to $1 \times 10^{-24}$
N/$\sqrt{\text{Hz}}$ \cite{Gierling2011}. Whilst, trapped-ions have
demonstrated force sensitivities below 500 $\times 10^{-24}$
N/$\sqrt{\text{Hz}}$ \cite{Biercuk2010}, with prospects of even lower
force sensitivities with novel geometries \cite{Maiwald2009}. On a
more macroscopic level, cantilever devices, are able to achieve force
sensitivities, reportedly down to $10^{-21}$ N/$\sqrt{\text{Hz}}$ and
$Q_m$-factors of greater than one million
\cite{Yasumura2000,Mamin2001,Rugar2004,Arcizet2006,Tao2014, Li2007,
  Moser2014, Weber2016}. Parallel to cantilever devices, toroidal
microresonators have achieved modest levels of force sensitivities
$\sim 1 \times 10^{-18}$ N$/\sqrt{\text{Hz}}$
\cite{Gavartin2012}. Such toroidal microresonators have achieved
$Q_m$-factors of up to $10^{9}$ \cite{Goryachev2012}, and position
sensitivities down to $1\times 10^{-19}$ m/$\sqrt{\text{Hz}}$
\cite{Schliesser2008}. These devices have a number of applications,
specially as on-chip force transducers \cite{Hu2013}. However, such
devices are strongly limited by noise due to mechanical coupling to
the environment.
\\
The fundamental requirements for a good force sensor are (according to
Eq. (\ref{eq:thforce})); good mechanical isolation from external noise
or a high $Q_m$-factor, low environmental temperatures and ideally low
oscillation frequencies.
\\
In levitated optomechanics, focused light is used for trapping
particles in air and vacuum \cite{ashkin1976}. Levitated particles are
more isolated mechanically from their environment than clamped systems
and exhibit high $Q_m$-factors of greater than $10^{6}$
\cite{Gieseler2012, Ranjit2015, vovrosh2016} in translational motion,
which, in principle, are limited only by the background gas pressure
and thus are predicted to reach $Q_m$ factors $> 10^{12}$.  Recently,
\citeauthor{Kuhn2017} \cite{Kuhn2017} have reached $Q_m$ of up to
$10^{11}$ for a driven rotational degree of motion of a levitated
nanorod at a few millibars of pressure, at room
temperature. Translational motion, generally, is calculated to have
force sensitivities of $1 \times 10^{-21}$N/$\sqrt{\text{Hz}}$
\cite{Gieseler12013}, whilst rotational or torsional degrees of
freedom of a trapped non-spherical nanoparticle are predicted to have
torsional force sensitivities of $2.4 \times 10^{-22}$
Nm/$\sqrt{\text{Hz}}$ \cite{Kuhn2017} to 2 $\times 10^{-29}$
Nm/$\sqrt{\text{Hz}}$ \cite{Hoang2016}. As a consequence of these
prospects, levitated optomechanics has attracted interest for
precision measurements in electron spin resonance
\cite{Hoang2016a,Neukirch2013}, short-range forces \cite{Geraci2010},
high-frequency gravitational waves \cite{Arvanitaki2013}, tests of
collapse models \cite{Bassi2013,bateman2014} and the
Schr\"odinger-Newton equation \cite{Großardt2015}, and direct dark
matter detection \cite{Bateman2014d}.
\\
Already, charged levitated particles have been studied in a hybrid
optical-electric Paul trap \cite{millen2015cavity}, in the search for
milli-charges \cite{Moore2014}, as well as, for demonstration of
charge control in nanoparticles \cite{Gierling2011, Frimmer2017}. The
control of charges on nanoparticles is essential for experiments to
prepare non-classical states of motion of the particle
\cite{bateman2014,Romero-Isart2011}. In addition, force detection at
$1.63 \times 10^{-18}$ N/$\sqrt{\text{Hz}}$ in levitated nanospheres
has already been demonstrated \cite{Ranjit2015} by experiment.
\par
Here, we take a detailed look at the interaction of an optically
levitated dielectric charged particle with an external electric field
as a case study for force sensing. We measure the effect of the
Coulomb interaction on the motion of a single nanoparticle, at high
vacuum ($10^{-5}$ mbar) by applying a DC and an AC electric field to a
metallic needle positioned near the trapped particle. These particles
can carry multiple elementary electric charges
($e = 1.6 \times 10^{-19}$ C), and we use the Coulomb interaction to
determine the number of elementary charges attached to the particle.
% --------------------------------------------------
%             Theory
% --------------------------------------------------
\par %new paragraph
The charge at the needle tip, $q_t$, for a given applied voltage is
according to Gauss's Law,
$\oint_s \mathbf{E} \cdot d\mathbf{s_t} = \frac{q_t}{\epsilon_0}$,
where $\mathbf{s_t}$ is the surface of the needle tip, $\epsilon_0$ is
the vacuum permittivity, and $\mathbf{E}$ is the electric field. The
electric field at any point in a potential, $V$, is given by
$-\nabla V=\mathbf{E}$. If, we approximate the needle tip as a sphere,
of radius, $r_t$, then $d\mathbf{s_t}=4\pi r_t d\mathbf{r}$. We get,
\begin{equation}\label{eq:tipcharge}
  \oint r_t \frac{dv}{dr}4\pi r_t dr = 4 \pi r_t V = \frac{q_t}{\epsilon_0}.
\end{equation}
We can then combine this with the Coulomb force acting on the particle
at distance, $d$,
\begin{equation}\label{eq:coulombforce}
  F_{C} = \frac{q_p q_t}{4 \pi \epsilon_0 d^2} = \frac{q_p V r_t}{d^2},
\end{equation}
where, $q_p$ is the charge on the nanoparticle. This additional force,
displaces the optically trapped particle.  With
Eq. (\ref{eq:coulombforce}) the nanoparticles' equation of motion can
be written as,
\begin{equation}\label{eq:forceEOM}
  \ddot{x}(t) + \Gamma_{0}\dot{x}(t) + \frac{k}{m}x(t) =
  \frac{F_{th}(t)}{m} + \frac{F_C}{m}e^{i\omega_{AC}t},
\end{equation}
where $k$ is the spring constant according to the optical gradient
force on the particle, and $F_{th}$ is a stochastic force originated
by a random process that satisfies the fluctuation-dissipation theorem
\cite{Kubo1966}. The AC driving frequency, $\omega_{AC}$, is zero when
considering the DC case. The time averaged position is now non-zero
relative to the trap centre and is given by,
\begin{equation}
    k_z \langle z\rangle = \cos(\theta)\frac{qVr_t}{\langle d\rangle^2},
    \label{eq:generalDCforce}
  \end{equation}
where $\langle z\rangle$ is the time average of the position in $z$ direction,
and $\theta$ is the angle between the direction of the force and the
$z$ direction. $d$ is the distance between the needle and
the trapped particle, thus we can write
$\langle z \rangle=d^{'} - \langle d \rangle$, where
$d^{\prime}$ is the distance between the centre of the laser focus and
needle tip. Taking $\langle z \rangle^2$ to be small in the resulting
quadratic equation and noting that $k_z = \omega_z^2m$, we get the
new average position to be,
\begin{equation} \label{eq:displacement}
  \langle z \rangle = \cos{\theta}\frac{q V r_t}{\omega_z^2md'^2}.
\end{equation}

For the AC case, $\omega_{AC} \neq 0$, and thus the AC contribution in
the equation of motion in Eq. (\ref{eq:forceEOM}) has to be considered.
Here we would like to look at the particles motion at the driving frequency,

\begin{equation}\label{eq:ACmotion}
  z(t) = z_0 e^{i\omega_{AC}t}.
\end{equation}

\begin{figure}
    \includegraphics[width=0.99\linewidth]{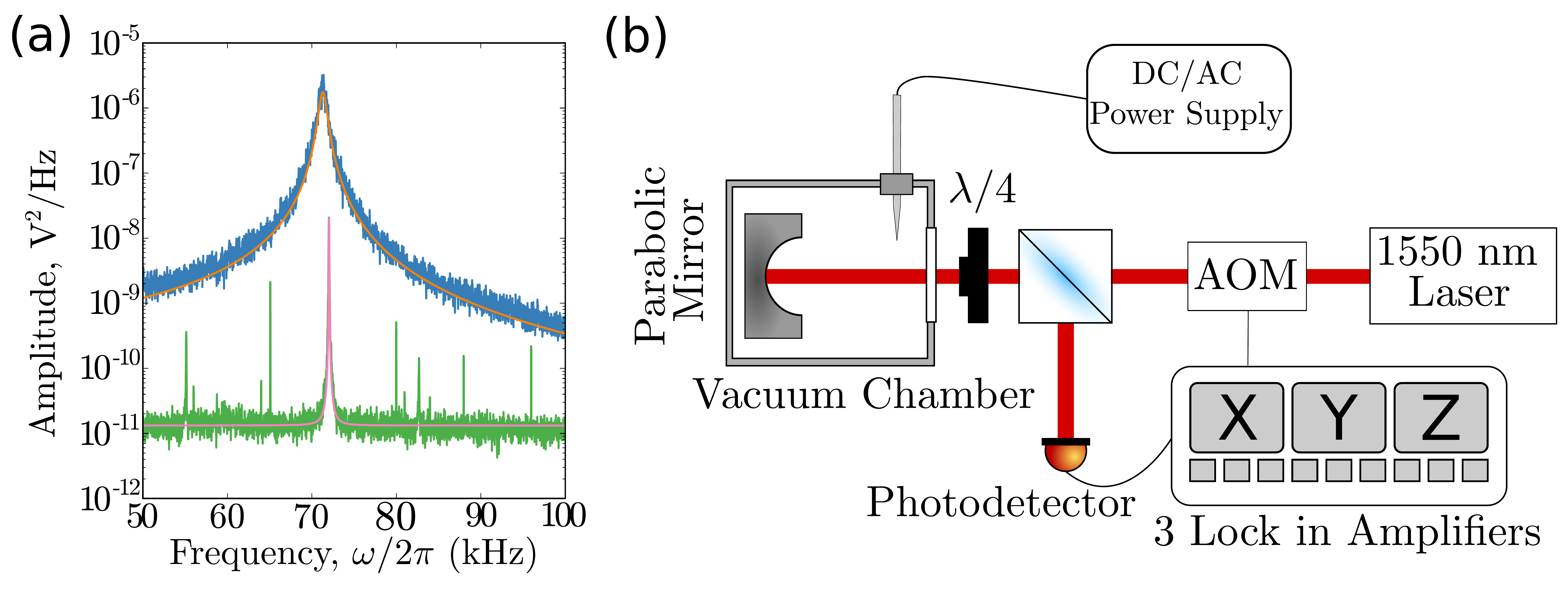}
    \caption{\textbf{Power Spectral Density and Experimental Setup}:
      \textbf{(a)} Cooling the motion in z-direction of a 41 $\pm$ 6
      nm radius particle, the upper spectrum (blue) is at $3\pm0.3$
      mbar and the lower spectrum is at $4.5 \times10^{-5}$ mbar. This
      corresponds to a temperature of $\sim$ 3 K from 300
      K. \textbf{(b)} A needle is connected via a high voltage vacuum
      feedthrough to either a DC power supply that can output up to 20
      kV or a signal generator for the AC experiments. The distance
      from the needle tip to the laser focus $d'$, is measured to be
      $39.6 \pm 0.8$ mm and at $\theta=45^{\circ}$. The mirror, along
      with the whole chamber, is earthed. }
    \label{fig:setupandcooling}
\end{figure}

Inserting Eq. (\ref{eq:ACmotion}) into Eq. (\ref{eq:forceEOM}) and
multiplying by the complex conjugate gives the peak height of the
particles motion at $\omega_{AC}$,
\begin{equation} \label{eq:ACefieldpsd}
  S_{AC}(\omega_{AC}) =
  \frac{1}{m^2}
  \frac{|F_{th}|^2+|F_{C}|^2}{((\omega_0^2-\omega_{AC}^2)^2+(\Gamma_0 + \delta\Gamma)^2\omega_{AC}^2},
\end{equation}
where $|F_{th}|^2=k_BT2m\Gamma_{0}/\pi$ and $\delta\Gamma$ is the
additional damping due to the parametric feedback. The analysis above
demonstrates that the DC contribution results in a shift in the
average position of the trap, whilst the AC driving introduces
resonance enhancement of the amplitude of the oscillation signal.
%
% --------------------------------------------------
%             Experimental Methods
% --------------------------------------------------
%
Without additional forces the Power Spectral Density (PSD) of the
particle's motion is given by,
\begin{equation}
    S_{xx}(\omega)=\gamma^2\frac{k_BT_0}{\pi m}
                \frac{\Gamma_0}{(\omega^2_0-\omega^2)^2+\Gamma_0^2\omega^2},
    \label{eq:psd}
\end{equation}
where $\gamma$ is the conversion factor that coverts the detection voltage to
metres \cite{vovrosh2016}. By fitting Eq. (\ref{eq:psd}) to the experimentally measured
PSD (see Fig. \ref{fig:setupandcooling}a), the damping from background
gas $\Gamma_0$ and feedback cooling $\delta\Gamma$, can be
determined. These fitted parameters can then be used to work out the
radius and mass of the particle, as well as, the centre-of-mass
(\emph{c.m}) temperature of the trapped particle from,
$T_{cm}=\frac{T_{0}\Gamma_{0}}{\Gamma_{0} + \delta\Gamma}$.

\par
In our experiments, we trap a silica nanoparticle (density,
$\rho_{Si02} \sim 2.65 $ g/cm$^3$), in a dipole trap. The optical
gradient force trap is realised using a 1550 nm laser and a high
numerical aperture (N.A.) parabolic mirror to produce a diffraction
limited focus.  The particle's position is measured by detecting the
interference between the light Rayleigh scattered by the particle and
the divergent reference light with a single photodiode (as shown in
Fig. \ref{fig:setupandcooling}b). The detected signal contains three
distinct frequencies for motion along $x, y, z$ directions, each of
which is sent to a lock-in amplifier. The amplifiers output to an
acoustic optical modulator (AOM) at twice the trap frequency with an
appropriate phase shift that counters the \emph{c.m} motion of the
particle, thus cooling the \emph{c.m} temperature. More details can be
found elsewhere \cite{vovrosh2016}. For both DC and AC case, we carry
out the experiments at a pressure of $1.6\times10^{-5}$ mbar and we
cool the particle motion to $\sim 3$ K in the $z$-axis (see
Fig.(\ref{fig:setupandcooling}a)). The needle that is used to generate
the DC/AC electric field is made of polished stainless steel and has a
tip radius of 100 $\mu$m. The distance between the trap centre to the
needle, $d'$, is measured to be $39.6 \pm 0.8$ mm and
$\theta=45^{\circ}$. To generate the DC field we connected the needle
to a high power supply (Berta High Voltage Power Supply 230 series)
and to generate the AC field we connect to a signal generator (TTi
TG1010A Programmable Function Generator).
\\
\begin{figure}%[htb] \centering
  \includegraphics[width=0.99\linewidth]{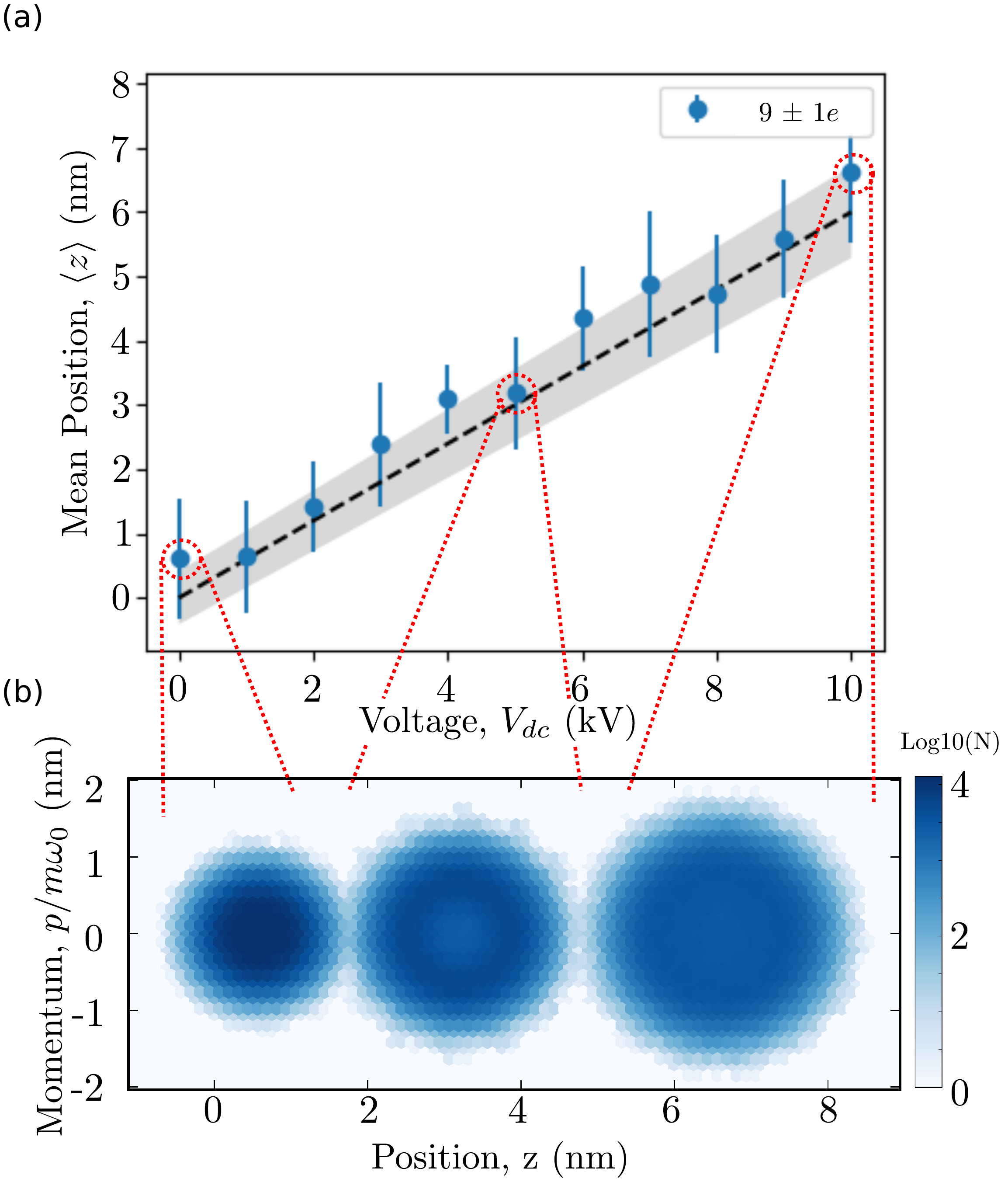}
  \caption{\textbf{Spatial displacement:} Data shows the spatial
    displacement in the $z$-direction for a 41 $\pm$ 6 nm particle for
    a DC voltage of 0-10 kV applied. \textbf{(a)} Displacement of the
    particle's mean position at the application of different DC fields
    produced by the needle. \textbf{(b)} Shows the displacement of the
    thermal state distribution at 0, 5, 10 kV to be 0.6 nm 3.1 nm and 6.6 nm,
    respectively. Using Eq (\ref{eq:displacement}) gives a charge of
    $9\pm 1e$. Through out these experiments the particles
    \emph{c.m} temperature is at $\sim 3$ K.}
  \label{fig:phasespace}
\end{figure}
\par
To study the effect of the DC field we levitate a $41 \pm 6$ nm (mass,
$m = 7.6 \times 10^{-19}$ kg) silica particle. The DC field generated
by the needle tip, modifies the effective potential experienced by the
particle. This modification leads to a shift in the mean position of
the particle. The spatial displacement is shown in
Fig. \ref{fig:phasespace}(b) and increases with increasing DC
voltage. Fig. (\ref{fig:phasespace}a) shows the displacement for
particle of charge of $ 9e \pm 1e$ and a spatial displacement of 6.6
nm for a $V_{dc}$ = 10 kV. The displacement operation increases the
\emph{c.m} temperature of the ensemble. The related heating can be
explained by the increasing absolute noise on the DC voltage. For
small displacements, such as those observed in the present study, the
trap stays harmonic. For voltages greater than 10 kV we often loose
the particle from the trap.
\begin{figure}
    \begin{overpic}[width=1\linewidth]{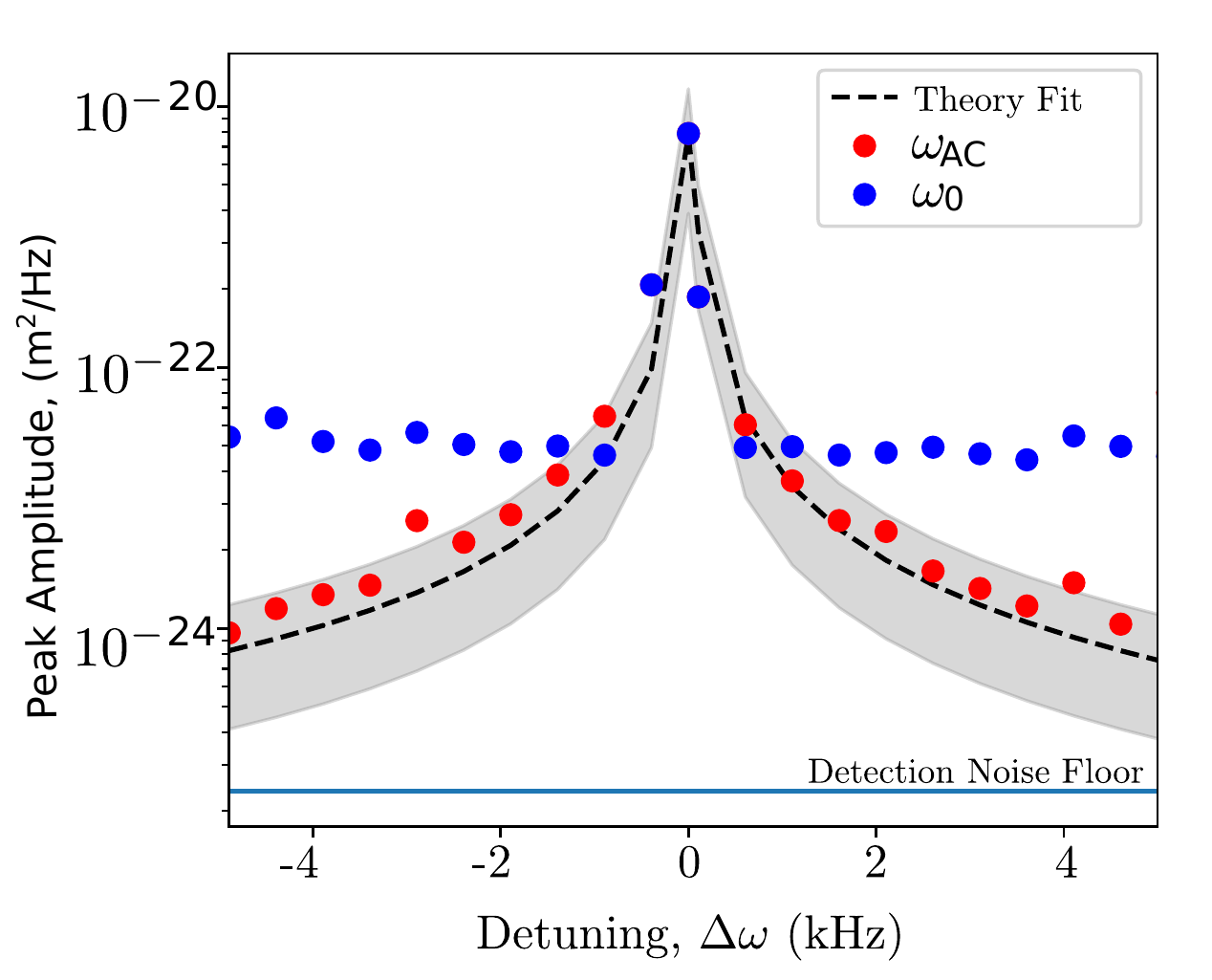}
    \end{overpic}
    % \begin{overpic}[width=0.49\linewidth]{zacexp} \put(12, 52){B}
    % \end{overpic}
    \caption[Cooling plot] {{\bf AC modulation: } Peak height of the
      driving AC field (in red) and levitated oscillator (in blue)
      frequencies. Eq. (\ref{eq:ACefieldpsd}) is fitted to the driving
      amplitude and the shaded region is the fitting error.  The
      detuning, $\Delta\omega = \omega_0 - \omega_{AC}$, is swept from
      low to high frequencies in steps of 500 Hz to show the full
      spectrum response. The averaged experimental noise floor is
      shown at 1.6 $\times 10^{-5}$ mbar. An AC force of
      \textbf{$3.0 \pm 1.5 \times10^{-20}$} N and a thermal force of
      $3.2\times10^{-20}$ N was measured with 10 second integration
      time. The particle has a radius of $50\pm6$ nm giving a mass of
      $1.4\times10^{-18}$ kg and a charge of $4\pm 3e$. }
	\label{fig:ACcoulomb}
\end{figure}
\par
In the case of the AC field, it is apparent from
Eq. (\ref{eq:ACefieldpsd}) that when $\omega_{AC}$ is far from
$\omega_{0}$ then the PSD signal is weak, however, as the two converge
there is a strong signal enhancement allowing much smaller forces to
be detected. Fig.(\ref{fig:ACcoulomb}) shows the peak heights, both
for the theory and experimental plots, demonstrating this enhancement
effect for a particle of radius $50 \pm 6$ nm. Using a pure sine wave
as the driving frequency, the detuning,
$\Delta\omega = (\omega_0-\omega_{AC})$, is swept in increments of 500
Hz across $\omega_0$. By fitting the recorded signal amplitude of the
driving field in Fig. \ref{fig:ACcoulomb} with
Eq. (\ref{eq:ACefieldpsd}), we obtain $F_{AC}$, which we measure, for
1 V amplitude of the AC field on resonance, to be
$3.0 \pm 1.5 \times 10^{-20}$ N integrated over 10 seconds. This
approaches a force sensitivity of $3.2 \times 10 ^{-20}$
N/$\sqrt{\text{Hz}}$, which is limited by gas collisions at the
pressure in the vacuum chamber.  Since, we obtain $F_{AC}$
experimentally, we can relate this to the number of charges on the
particle as,
\begin{equation} \label{eq:ACForce}
  F_{AC} = \frac{q_{p}Vr_{t}}{d^{'2}}.
\end{equation}
Thus, the number of elementary charges on the nanoparticle, in the AC
experiment, is calculated to be $4\pm 3e$. The resonant driven signal
is enhanced by a factor of 200 compared to the undriven system.
% --------------------------------------------------
%             Discussions
% --------------------------------------------------
\par
The limiting factor to reach even lower force sensitivities than
$10^{-20}$ N/$\sqrt{\text{Hz}}$ can be associated to noise in the
present system. In general, this noise is a summation of detector
noise, electronic noise via the feedback system, mechanical noise in
the optical elements, classical thermal noise due to gas collisions
and finally, the standard quantum limit (SQL). The dominating noise
for the current system is the thermal noise floor according to
background gas collisions at a pressure of $10^{-5}$ mbar. In
addition, long term laser power drifts of approximately 1\%, at
timescales of hours, are observed. This is due to thermal drifts in
the fibre optics which consequently causes a change in polarisation,
which effects the trapping power and thus introduces drifts in the
trapping frequencies. This power drift predominantly effects the DC
experiments, which requires the measurement of many different DC
voltages and takes many hours to carry out. In additional, at short
time scales, we also have electronic noise due to the feedback system
\cite{vovrosh2016}. The corresponding averaged experimental noise
floor is shown Fig. \ref{fig:ACcoulomb}. The error bars on the
particle size, mass and charge are dominated by the uncertainty in the
pressure readings, which is accurate to $15$\%.
\\
The classical thermal noise, which has already been discussed in Eq.
(\ref{eq:thforce}) and for levitated systems is physically realised by
gas collisions, puts a strong limit on the systems sensitivity. But
with modification to the current setup, i.e.  for lower pressure
($\sim 10^{-9}$ mbar) and with a smaller particle (r $\sim 10$ nm) and
trapping frequencies of $\sim$100 kHz, force sensitivities down to
$1 \times 10^{-24}$ N/$\sqrt{\text{Hz}}$ can be reached.
\\
At this limit of $1 \times 10^{-24}$ N/$\sqrt{\text{Hz}}$, it is
envisaged that the standard quantum limit (SQL) for the system would
be reached. The SQL, which can be written as
$S_{FF}^{SQL} = \sqrt{\hbar \omega_0 m/2\tau_F} $, where $\tau_F$ is
the rate of the measurement carried out on the particle
\cite{Braginsky1995}, is calculated for the current system to be
$6 \times 10^{-24}$ N$/\sqrt{\text{Hz}}$.
\par
In conclusion, we have measured the response of an optically levitated
charged nanoparticle to a DC and an AC electric field. We have
observed spatial displacement of the centre of the thermal motional
state of the particle in phase space by up to 6.6 nm for an applied DC
field of 10 kV. We find that by applying an AC field amplitude of 1 V
on resonance we are able to measure a force of $3.0 \times 10^{-20}$
N. The sensitivity can, in future experiments, be improved by lowering
the noise floor, which is limited by the thermal noise of gas
collisions at $10^{-5 }$ mbar. We extrapolate that by optimising
particle size, pressure and mechanical frequency we can reach
SQL. Then, techniques such as position or momentum squeezing of
mechanical oscillators \cite{Wollman2015, Rashid2016, Riedinger2016}
may be used to increase for sensitivities even further. While this
gives a direct perspective for the use of levitated optomechanics for
force sensing applications, the system is also suitable for
fundamental physics problems. The experiment can be used for a
non-interferometric test of the quantum superposition principle
\cite{Bassi2013}. Specifically, the continuous spontaneous
localization (CSL) model \cite{Ghirardi1990}, which gives a
quantitative violation of the superposition principle, predicts a
slight increase in temperature of the trapped nanoparticle. This
effect, as discussed in \cite{Vinante2016,vinante2016upper}, can be
used to set bounds \footnote{The ability of the experimental setup to
  set bounds on the CSL parameters is related to the precision
  $\Delta T$ with which we can measure the temperature $T$ of the
  system. Specifically, we have estimated the bounds on the CSL
  parameters from the condition
  $\Delta T_{\text{CSL}}(\lambda,r_{C})\geq\Delta T$, where
  $\Delta T_{\text{CSL}}(\lambda,r_{C})$ denotes the theoretical
  temperature increase predicted by the CSL model.} on the CSL
parameters, namely, on the localization rate $\lambda$ and on the
localization length $r_{C}$. The minimum value of $\lambda$ that could
be excluded by the current experimental setup is
$\approx10^{-6}s^{-1}$ (achieved at
$r_{C}\approx0.3\times10^{-7}\text{m}$), which corresponds to a
macroscopicity measure \cite{Nimmrichter2013} of
$\mu\approx12$. Increasing the size of the trapped particle to $R=300$
nm, which can be trapped by the current experimental setup, would
improve the bounds on $\lambda$ by two orders of magnitude.
\par
{\it Acknowledgments:} We thank Phil Connell and Gareth Savage for
expert technical help during the realisation of the experimental
setup. We also like to thank Chris Timberlake, Markus Rademacher,
Ashley Setter for the QPlots package and discussions. We wish to thank
for funding, The Leverhulme Trust and the Foundational Questions
Institute (FQXi).

\bibliography{main.bib}
\bibliographystyle{apsrev4-1}
\end{document}